\documentclass[12pt,a4paper]{article}
\usepackage[dvips]{epsfig}
\usepackage[english]{babel}
\usepackage{amsmath}
\usepackage{amssymb}

\begin{document}

\thispagestyle{empty}

\title{The two-angle model and the phase diagram for Chromatin}

\author{Philipp M. Diesinger and Dieter W. Heermann
         \\ {} \\
           Institut f\"ur theoretische Physik \\
           Universit\"at Heidelberg \\
           Philosophenweg 19 \\
           D-69120 Heidelberg\\
           and\\
           Interdisziplin\"ares Zentrum\\
           f\"ur Wissenschaftliches Rechnen\\
           der Universit\"at Heidelberg \\ {} \\
\vspace {3ex}}

\newpage

\maketitle
\vfill\eject
\begin{abstract}
We have studied the phase diagram for chromatin within the framework of the two-angle model. Rather than improving existing models with finer details our main focus of the work is getting mathematically rigorous results on the
structure, especially on the excluded volume effects and the effects on the energy due to the long-range forces and their screening. Thus we
present a phase diagram for the allowed conformations and the Coulomb energies.
\end{abstract}

\vspace{2cm} {\bf Keywords: chromatin, phase diagram, exact results}

\vfill\eject

\section{Introduction}

Each protein aggregate together with its wrapped DNA comprises a nucleosome core particle with a
radius of about 5nm and a height of about 6nm; with its linker DNA it is the fundamental chromatin
repeating unit. It carries a large electrostatic charge \cite{Khraounov}. Whereas the
structure of the core particle has been resolved up to high atomic resolution \cite{Luger}, there is still
considerable controversy about the nature of the higher-order structures to which they give rise. When stretched the chromatin string appears to look like "beads-on-a-string" in electron micrographs.

The recent accumulation of evidence that the chromatin structure above the level of the core particle
plays a key role in determining the transcriptional status of genes and genetic loci \cite{Fletcher,Felsenfeld}
illustrates the critical importance of understanding the fundamental folding properties of nucleosome
arrays: Studies of chromatin compaction in response to changes in the ionic environment \cite{Widom86}
have established that the phenomenon can be accounted for by electrostatic interactions between DNA,
histone proteins and free ions \cite{clark}. Major contributions to these interactions are provided by the
N-terminal domains of the core histones, which contain roughly half of the basic amino acids of the
octamer, and the C-terminal domains of the linker histones, which contain about $\frac{3}{5}$ of the positive
charges in these molecules. Indeed, chromatin compaction requires the presence of the core histone N
termini and 30nm chromatin fibers are not formed in the absence of linker histones. However, the precise
interactions that lead to specific chromatin higher-order structures have remained elusive.

The beads-on-a-string structure can be seen clearly when chromatin is exposed to very low salt concentrations, and is known as the 10-nm-fiber,
since the diameter of the core particle is about 10nm. With increasing salt concentration, i.e. heading towards physiological conditions ($c
\approx$100mM), this fiber appears to thicken, attaining a diameter of 30nm. The absence of the extra linker histones (H1 or H5) leads to more
open structures; so it is surmised that the linker histones act near the entry-exit point of the DNA; they carry an overall positive charge and
bind the two strands together leading to a stem formation \cite{Bednar,Bednar2,Stasiak}. Increasing the salt concentration decreases the
entry-exit angle $\alpha$ of the stem as it reduces the electrostatic repulsion between two strands.

Longstanding controversy \cite{van Holde,Widom89,Zlatanova} surrounds the structure of the 30nm fiber, for which there are mainly two competing
classes of models: the solenoid models \cite{Finch,Thoma} and the crossed-linker-models \cite{Woodcock,Horowitz,Leuba}. In the solenoid model it
is assumed that the chain of nucleosomes forms a helical structure with the axis of the core particles being perpendicular to the solenoidal
axis (the axis of an octamer corresponds to the axis of the superhelical path of the DNA that wraps around it). The DNA entry-exit side faces
inward towards the axis of the solenoid. The linker DNA is required to be bent in order to connect neighboring nucleosomes in the solenoid. The
other class of models assumes straight linkers that connect
nucleosomes located on $opposite$ sides of the fiber. This results in
a three-dimensional crossed-linker-pattern. Such an arrangement with peripherally arranged nucleosomes and internal linker DNA segments is fully
consistent with observations in intact nuclei and also allows dramatic changes in compaction level to occur without a concomitant change in
topology.

Images obtained by electron cryomicroscopy (EC-M) should in principle be able to distinguish between the structural features proposed by the
different models mentioned above. The micrographs show a crossed-linker-pattern at lower salt concentrations and they indicate that the
chromatin fiber becomes more and more compact when the ionic strength is raised towards the physiological value \cite{Bednar}. However, for these
denser fibers it is still not possible to detect the exact linker geometry. Experiments on dinucleosomes have been made to check whether the
nucleosomes collapse upon an increase in ionic strength. A collapse would only occur if the linker bends, and an observation of this phenomenon
would support the solenoid model. The experiments by Yao et al. as well as more recent experiments by Butler and Thomas indeed reported a
bending of the linkers but do not agree with experiments by Bednar et al. and by others that did not find any evidence for a collapse.

In our work we shall follow the ideas put forward by Woodcock et al.~\cite{Woodcock}
and Schiessel et al.~\cite{Schiessel} but follow a rigorous mathematical approach.

\section{ The two-angle-model}

\begin{figure}[ht]
\begin{center}
\includegraphics[width=\textwidth]{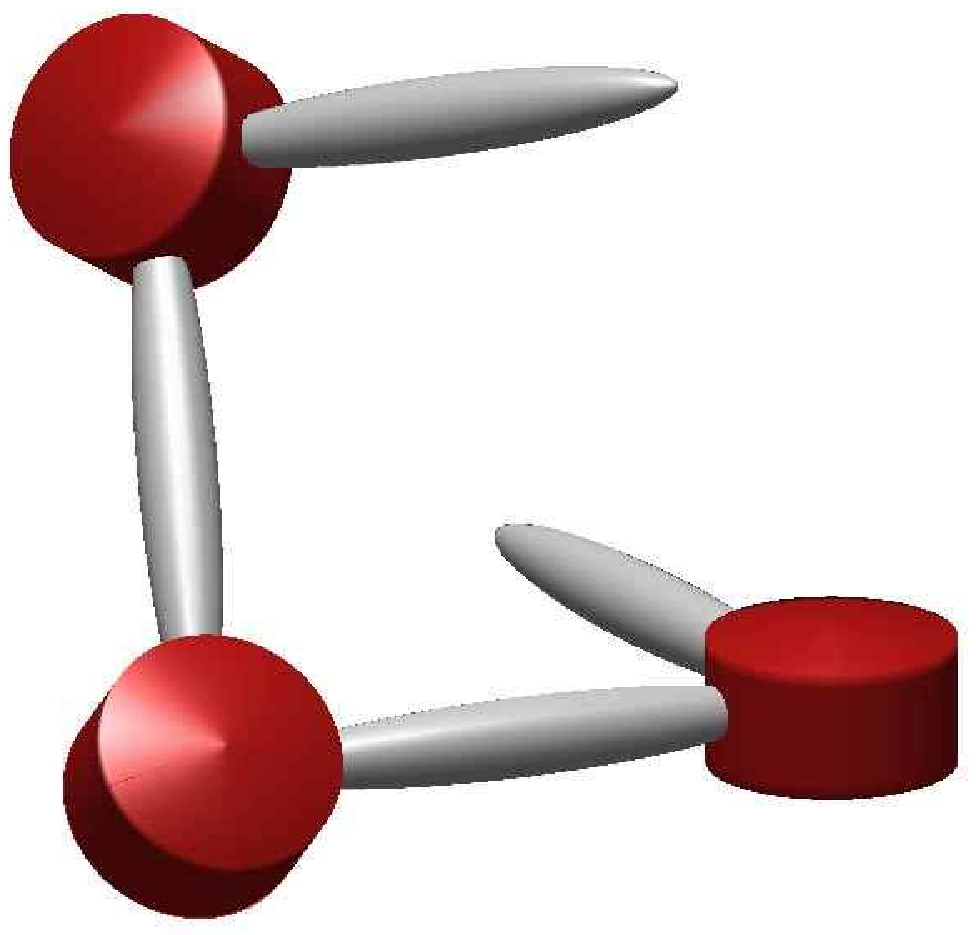}
\caption{\label{fig:basic_def} Basic definitions of the two-angle model : The entry-exit angle $\alpha$, the linker length b, and the rotational
angle $\beta$.}
\end{center}
\end{figure}

Following Woodcock~\cite{Woodcock} et al. and Schiessel et al.~\cite{Schiessel} we
consider four consecutive nucleosomes (cf. Fig. \ref{fig:basic_def}): $N_0$,$N_1$,$N_2$ and $N_3$ $\epsilon \; \mathbb{R}^3$ within the chain.
$N_3$ is a function of $N_0,..,N_2$ by fulfilling the following conditions:

\begin{align*}
&i) \;\;\;\sphericalangle \left ( (N_0-N_1),(N_2-N_1) \right) = \alpha; \\&ii)\;\;  \| N_2-N_1\| = b_2,\| N_0-N_1\| = b_1,\| N_3-N_2\| = b_3,\;
\text{with} \; b_1,...,b_3=b;\\ &iii)\;  P := \{ r \;  \epsilon \; \mathbb{R}^3 \mid \exists \; \lambda , \; \mu \; \epsilon \; \mathbb{R},\;
\text{such that } r = N_1 + \lambda (N_0-N_1) + \mu (N_2-N_1)  \} \\ & \;\;\;\;\;P' := \{ r \; \epsilon \; \mathbb{R}^3 \mid \; \exists \lambda
' , \; \mu ' \; \epsilon \; \mathbb{R}, \text{ such that } r = N_1 + \lambda '(N_2-N_1) + \mu ' (N_3-N_1) \} \\ & \;\;\;\;\;
\sphericalangle(P,P')=\beta.
\end{align*}

By straightforward considerations this leads to the following expression for $N_3$:

\begin{align*}
& N_3 =\mathfrak{R}_{\beta}^{\hat{w}} \; \mathfrak{R}_{\pi - \alpha}^{\hat{v}} \;  \left ( N_2 + b_3 \cdot \frac{(N_2-N_1)}{\| N_2 - N_1 \|}
\right ) \\ \\ &\hat{v} := \frac{(N_2-N_1)\times(N_0-N_1)}{\|(N_2-N_1) \times (N_0-N_1)\|}; \quad \hat{w} := \frac{N_1-N_2}{\|N_1-N_2\|};
\end{align*}
where $\| N_2 - N_1\| = b_2$ and $\mathfrak{R}_{\varphi}^{v}$ is the orthogonal rotational transformation matrix defined by the axis $v \;
\epsilon \; \mathbb{R}^3$ and the rotation angle $\varphi \; \epsilon \; [0,2\pi]$ (with respect to the right-hand-rule). (Note that the
chromatin fibers described by these formula do not show a tangential distance between the ingoing and the outcoming DNA linkers).

So the geometrical structure of the necklace is determined entirely by $\alpha$, $\beta$ and $b$. But this model only describes the linker
geometry and does not account for any forms of nucleosome-nucleosome interaction. Furthermore it assumes $straight$ linkers. It is still not
completely clear whether the linkers of the 30nm fiber at high salt concentration are straight. But the bending of the linkers would cost about
10$kT$ and straight linkers are very likely, cf. \cite{Bednar}.

For every ideal chromatin fiber with a certain set of values ($\alpha$, $\beta$, $b$) it is possible to construct a spiral with radius $R$ and a
gradient $m$ so that all the nucleosomes are located on this spiral (cf. \cite{Schiessel2}). The nucleosomes are placed along the spiral in such
a way that successive nucleosomes have a fixed (Euclidean) distance $b$ from one another (in fact there are many such spirals, but the
interesting spiral is the one with the largest gradient $m$). The parametrization of the spiral is given by

\begin{align*}
\gamma(t) := \begin{pmatrix} R \cdot cos(\frac{a \cdot t}{R})& \\R \cdot sin(\frac{a \cdot t}{R})& \\t& \end{pmatrix} , \; t \; \epsilon \;
\mathbb{R}\end{align*}
where $R$ is the radius and $m=\frac{1}{a}$ is the gradient of this master solenoid (which follows from $\dot{r}(0)=(0,a,1)^T$ ).

By straightforward calculations one can show the following relation:
\begin{equation}\label{eq:b^2} b^2=2R^2 \left (1-\cos \left (\frac{a \cdot d}{R} \right ) \right ) + d^2 \quad .\end{equation}

Furthermore, $\cos(\pi-\alpha)=\frac{<r_0 \mid r_2>}{\|r_0^2\|}$ with $r_0 = R_1 -R_0$, $r_0 = R_2 -R_1$ and $r_2 = R_0 - R_{-1}$ leads to:

\begin{equation}\label{eq:alpha}
\cos(\pi-\alpha)=\frac{2R^2\cos \left (\frac{a \cdot d}{R} \right ) \left (1-\cos \left (\frac{a \cdot d}{R} \right ) \right )+d^2}{2R^2 \left
(1-\cos \left (\frac{a \cdot d}{R} \right ) \right )+ d^2 } \quad .
\end{equation}

Finally, $\beta$ can be calculated by evaluating $\cos(\beta)= \frac{<r_0 \times r_1 |r_2 \times r_0>}{\| r_0 \times r_1\| \cdot \| r_2 \times
r_0 \|}$, as the angle between two successive planes (cf. iii):
\begin{equation}\label{eq:beta}
\cos(\beta)=\frac{d^2 \cos \left (\frac{a \cdot d}{R} \right ) + R^2 \sin^2 \left (\frac{a \cdot d}{R} \right )}{d^2+R^2 \sin^2 \left (\frac{a
\cdot d}{R} \right )} \quad .
\end{equation}

So equations \ref{eq:b^2}, \ref{eq:alpha} and \ref{eq:beta} relate $R$, $a$ and $d$ to $\alpha$, $\beta$ and
$b$ and thus the \emph{global} fiber geometry to these \emph{local} variables. These equations are
important for further calculations.

The inverse transformation, which relates the local fiber properties to the global spiral geometry (i.e. ($ \alpha, \beta,b$) $\Rightarrow$
($R,m=\frac{1}{a},d$)) is very useful, too, and given by (cf. \cite{Schiessel2}):
\begin{equation}\label{eq:Radius of the Spiral}
R = \frac{b \cdot \cos \left (\frac{\alpha}{2} \right )}{2 \left (1-\sin^2 \left (\frac{\alpha}{2} \right)\cos^2 \left (\frac{\beta}{2} \right )
\right )}\end{equation}\begin{equation}
m=\frac{1}{a}=\frac{2 \sin \left ( \frac{\beta}{2} \right ) \sqrt{1- \sin^2 \left (\frac{\alpha}{2} \right ) \cos^2 \left (\frac{\beta}{2}
\right )}}{\cot \left ( \frac{\alpha}{2} \right ) \arccos \left (2 \sin^2 \left (\frac{\alpha}{2} \right ) \cos^2 \left (\frac{\beta}{2} \right
) - 1  \right ) }\end{equation}
\begin{equation}\label{eq:d of the spiral}
d=\frac{b \sin \left ( \frac{\beta}{2} \right )}{\sqrt{\csc^2 \left (\frac{\alpha}{2} \right ) - \cos^2 \left ( \frac{\beta}{2} \right ) } }
\overset{\beta \ll 1}{\approx} \frac{b \cdot \beta}{2\sqrt{\csc^2(\frac{\alpha}{2})-1}}+\;o(\beta)^3 \quad .
\end{equation}These three equations relate the local fiber geometry to the global properties of the associated spiral.

\section{Phase Diagram}

\begin{figure}[ht]
\begin{center}\hspace{0cm}
\includegraphics[width=\textwidth]{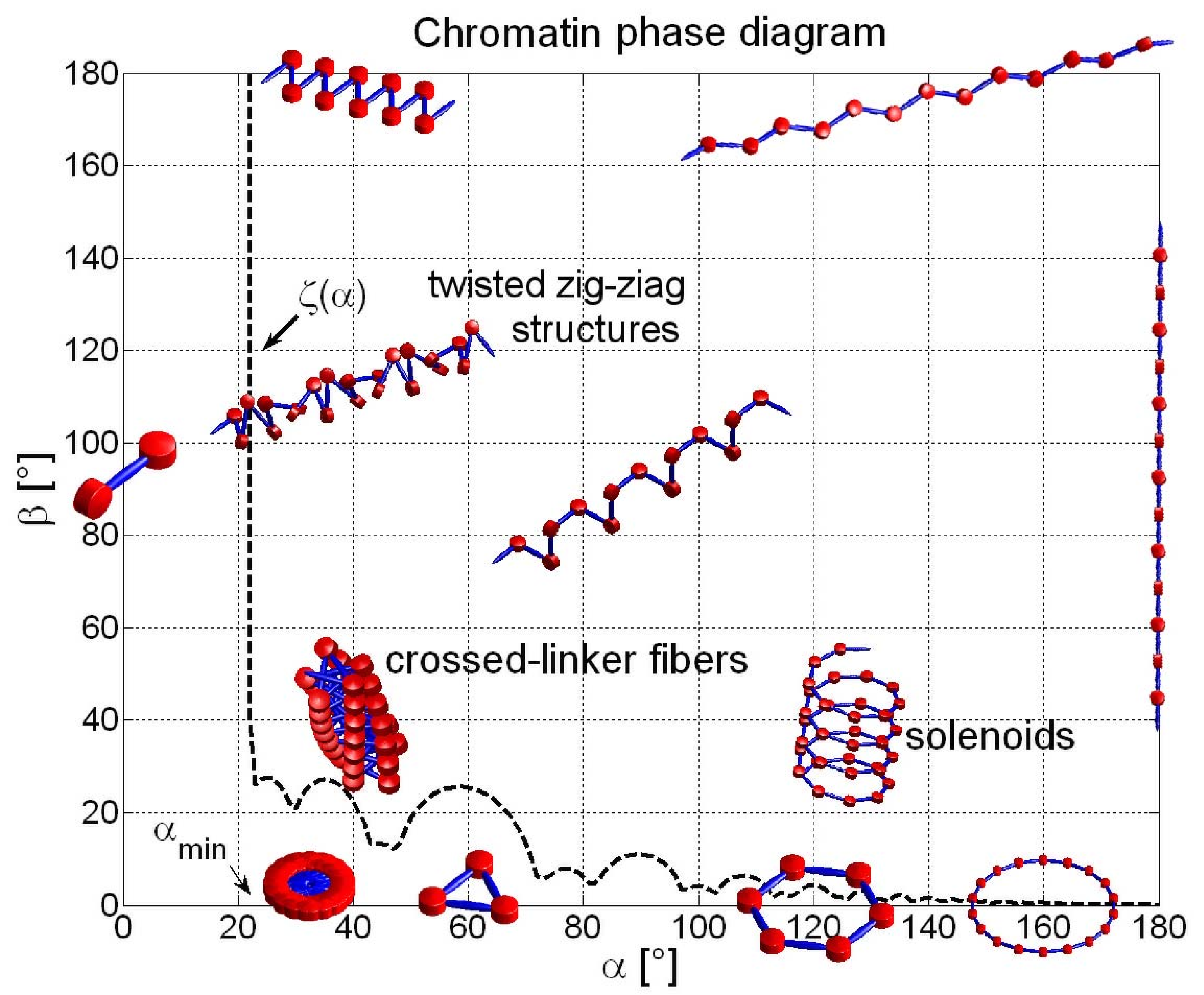}
\caption{\label{fig:phase_diagram} The chromatin phase diagram with some chromatin structures for different values of $\alpha$ and $\beta$. The
solenoid and crossed-linker structures are most important. The dotted line is the function $\zeta$($\alpha$) which is explained in the text.}
\end{center}
\end{figure}

Fig. \ref{fig:phase_diagram} shows the chromatin phase diagram: For every point \begin{equation*}\gamma_n\; \epsilon \; \Gamma_n :=
\{(a_i)_{i=1...n}, \; a_i \; \in  \mathbb{R}^3\;  \mid \text{ the series $(a_i)$ fulfills i)-iii) with } \alpha, \; \beta \; \epsilon \;
[0,\pi]\}\end{equation*} there is a certain fiber structure (i.e. $\Gamma_n$ is the phase space of all ideal fibers of length $n$). These
structures will be discussed in the following two sections. They can be divided into planar and three-dimensional structures.$\\$ In fact the
true phase diagram would be at least three-dimensional, because the fiber's length $N$ is an important parameter, too. It plays an important
role from the numerical point, when considering different functions, which depend on the fiber's geometry as well as on its length such as
nucleosome-density or excluded-volume-interactions. In these cases the fiber length should be as small as possible to lower the running time of
the program but it should also be large enough to obtain the right physical insights. This optimization principle is not always easy to fulfill -
especially in the latter case of the hard-core excluded-volume-interactions.$\\$ Another parameter which is not shown in Fig.
\ref{fig:phase_diagram} is the linker length $b$. But this can be easily ignored as the linker length is considered to be a constant for ideal
fibers, and the mean value of it is fixed.$\\$ The classical solenoid model of Finch and Klug (cf. \cite{Finch}) is found in the large $\alpha$
and small $\beta$ area in Fig. \ref{fig:phase_diagram} (in their case the linkers were bent). Various other structures with $\alpha$=30° and
different values of $\beta$ were found by Woodcock et al. (cf. \cite{Woodcock}). These are located on the left side of Fig.
\ref{fig:phase_diagram}. So all these possible 30nm-structures can be found in the phase diagram.

If either one of the angles $\alpha$ or $\beta$ equals 0 or $\pi$ the resulting structure will be planar. Although it seems that the planar
structures are not of much interest from the physical point of view it turned out that they play a very important role for the excluded volume
structure of the phase diagram, and therefore they will be discussed here in detail.$\\$ In the case $\beta=\pi$ and $\alpha$ arbitrary the
fiber forms 2D zig-zag-like structures, as shown at the top of Fig. \ref{fig:phase_diagram}. The length of a fiber consisting of N monomers is
given by $L = d \cdot N = b \cdot \sin \left (\frac{\alpha}{2}\right)N$ (cf. Eq. \ref{eq:d of the spiral}) and the diameter is given by $ D = b
\cos \left (\frac{\alpha}{2}\right)$ (cf. Eq. \ref{eq:Radius of the Spiral}). The length of the fiber increases with increasing $\alpha$. In the
case of $\alpha = \pi$ with an arbitrary value of $\beta$ the fiber is a straight line and for $\alpha = 0$ and $\beta \; \epsilon \; $[0,$\pi$]
it corresponds to linkers that go back and forth between two positions (cf. Fig. \ref{fig:phase_diagram}). $\\$Consider now the important case
of the line $\beta=0$ at the bottom of Fig. \ref{fig:phase_diagram}.$\\$ If $\alpha\approx\pi$, the fiber forms a circle (with radius $R \approx
\frac{b}{\pi-\alpha}$ as follows from Eq. \ref{eq:Radius of the Spiral}). Its radius converges towards infinity for $\alpha \rightarrow \pi$.
For certain values of $\alpha$ the fiber forms a \emph{regular polygon}. For instance the value $\alpha = \frac{\pi}{2}$ corresponds to the
square and $\alpha = 60°$ is the regular triangle (cf. Fig. \ref{fig:phase_diagram}). To characterize these regular polygons one needs two
variables: At first the number of the tips $i$ and secondly the order $n$ of the polygon which gives the number of loops the fiber needs to
arrive at the starting point again.

These special values of $\alpha = \alpha_i^n$ are
given by

\begin{equation}\label{eq:alpha_i^n} \alpha_i^n = \pi - \left ( \frac{n \cdot 2 \pi}{i} \right ) \text{, with } i,n \; \epsilon \; \mathbb{N} \text{ and } i \geq 2n\text{, such that  }
 \end{equation} \begin{align*}
\nexists \; n', i' \; \epsilon \; \mathbb{N} \text{ with } n'< n \text{ and } \alpha_i^n=\alpha_{i'}^{n'}\;\;\; (\vartriangle).\end{align*}

The order $n$ of $\alpha_i^n$ is a measure of its influence on the forbidden surfaces of the excluded volume structure of the
phase diagram. Therefore all different values of $\alpha_i^n$ have to be sorted by $n$ first and secondly by $i$ (if $n_1 > n_2$, then
$\alpha_i^{n_1}$ is "more important" than $\alpha_i^{n_2}$ and if $i_1>i_2$ then $\alpha_{i_1}^n$ is "more important" than $\alpha_{i_2}^n$) -
this is what makes $(\triangle)$ necessary: For example $\alpha_3^1=\alpha_6^2=\alpha_9^3$ but the order of these three $\alpha_i^n$ is always
$n=1$ and therefore they are all in the same equivalence class (the special case $i=2n$ leads to one-dimensional structures with $\alpha_2^1=0$
- this value of $\alpha_i^n$ of highest order in $n$ and $i$ plays an important role for the forbidden area in the phase diagram and is
therefore mentioned here, too).$\\$ All $\alpha_i^n$ are irrational numbers (cf. Eq. \ref{eq:alpha_i^n}). For numerical reasons it is therefore
useful to give all angles in degrees, because these are always \emph{rational} numbers.

Furthermore, $ (\triangle) \; \Leftrightarrow \;$\emph{ $\; \nexists \; \alpha,\beta,b \in \mathbb{N}$ with $n=\alpha \cdot b$ and $i=\beta\cdot
b$, which means that $n$ and $i$ are coprime} $(\bigstar)$ \;\; as $\alpha_i^n$ depends only on the relation $\frac{n}{i}$. $\\$ "$\Rightarrow$"
Let $n$ be an arbitrary element of $\mathbb{N}$. Now assume that $i$  $\in \mathbb{N}$, $i\geq 2n$ and $n$ and $i$ have a common divisor $b$
i.e. $ \exists \; \alpha , \; \beta \; \text{and} \; b \in \mathbb{N}$ such that $i=\alpha \cdot b$ and $n=\beta \cdot b \; \Rightarrow \;
\frac{n}{i}=\frac{\beta \cdot b}{\alpha \cdot b} =\frac{\beta}{\alpha}=\frac{\frac{n}{b}}{\frac{i}{b}}$, furthermore $i\geq 2n$ indicates
$\frac{i}{b}\geq 2 \frac{n}{b} \; \Rightarrow \; \frac{\beta}{\alpha}\leq \frac{1}{2} \text{ which means } \exists \; n', i' \; \epsilon \;
\mathbb{N} \text{ with } i'\geq2n' \text{ and } \frac{n'}{i'}= \frac{n}{i} \text{, namely } n' = \beta \text{ and } i' = \alpha$. This means
$\neg \bigstar$ implies $\neg \triangle$ which is equivalent to $\triangle \Rightarrow \bigstar$.$\\$ "$\Leftarrow$" To prove the other
implication above it suffices to show that if $n$ and $i$ are coprime, there are no numbers $m$, $j$ with $m < n$ such that $nj = mi$. Dividing
this equation by gcd$(n,m)$ one gets a new equation of the same structure with $m'$ and $n'$ instead of $m$ and $n$, fulfilling gcd$(m', n') =
1$. Moreover, gcd$(n,i)=1$ leads to gcd$(n',i)=1$. Thus gcd$(m'i, n') = 1$ $\Rightarrow$ $j = 1$ which contradicts to $j>2n$.

Now consider the case that $n$ and $n'$ are coprime with $n'>n$, $i \geq 2n$ and $i'\geq 2n'$ such that $\frac{n}{i}$, $\frac{n'}{i'} \not \in
\frac{1}{\mathbb{N}}$ (which means that they are not equivalent to any order-1-value of $\alpha_i^n$). Furthermore, assume that
$\frac{n'}{i'}=\frac{n}{i} \Rightarrow \; i' = \left ( \frac{n'}{n} \right ) \cdot i$, it is clear that $\left ( \frac{n'}{n} \right ) \not \in
\mathbb{N}$ and $i \in \mathbb{N}$, therefore $\exists \; \alpha \in \mathbb{N}$ so that $i=\alpha \cdot n$ which contradicts $\left (
\frac{i}{n} \right ) \not \in \mathbb{N}$. This means that if two orders $n_1$ and $n_2$ are coprime with $n_1, n_2 > 1$ , there are
\emph{never} equal values of $\alpha_i^{n_1}$ and $\alpha_{i'}^{n_2}$ for all possible $i \geq 2n$, $i' \geq 2n'$ and $i,i' \in \mathbb{N}$.
For example, $n$ and $n+1$ are always coprime numbers and therefore have never common $\alpha_i^n$.\\

So for a given order $n$ the possible values of $\alpha_i^n$ depend on the prime factor dismantling of $n$ and therefore are very irregular. If
$n$ is a prime number, all $i \geq 2n$ with $i \not \in n\cdot \mathbb{N}$ are allowed: for $n=2$ all $i \geq 4$ with $i\not \in2\cdot
\mathbb{N}$ are allowed and for $n=3$ all $i$ with $i \geq 2n$, $i \;\not \in\;3\cdot \mathbb{N}$ are allowed (and for $n$ = 1 all $i \geq 2$
are allowed). So between two values of $\alpha_i^1$ there is one of $\alpha_i^2$ and two of $\alpha_i^3$.

The distances $\Delta_{i_1}^n=\alpha_{i_2}^n-\alpha_{i_1}^n$ between two consecutive $i_2>i_1$ are given below for the first three and
additional for all prime orders:\begin{align*}
&n=1\text{ : } \alpha_i^1 = \pi - \frac{2\pi}{i}, \; \left ( i \geq 2n \right )\; \Rightarrow \Delta_i^1= \frac{2\pi}{i(i+1)}\\
&n=2\text{ : } \alpha_i^2 = \pi - \frac{4\pi}{i}, \; \left ( i \geq 2n \text{ and }i \not \in \; 2\cdot \mathbb{N} \right )\;
\Rightarrow \Delta_i^2= \frac{8\pi}{i(i+2)}\\
&n=3\text{ : } \alpha_i^3 = \pi - \frac{6\pi}{i}, \; \left ( i \geq 2n \text{ and }i \not \in \; 3\cdot \mathbb{N} \right )\;
\Rightarrow \Delta_i^3= \frac{\mod(i,3)\cdot6\pi}{i(i+\mod(i,3))}\\
&n \text{ prime} \text{ : } \alpha_i^n = \pi - \frac{n \cdot 2\pi}{i}, \; \left ( i \geq 2n \text{ and }i \not \in \; n\cdot \mathbb{N} \right )\;
\Rightarrow \Delta_i^n= \frac{m\cdot n \cdot 2\pi}{i(i+m)}\quad .\\
\end{align*}
This shows furthermore that $\Delta_i^n \overset{n\rightarrow\infty}{\rightarrow}0$, because the counter is always $\sim n $ and the leading
term of the denominator is at least $\sim n^2$ (as $i>2n$).

The most interesting structures of the chromatin phase diagram are the following two cases: solenoid-like structures and fibers with
crossed-linkers. An extensive discussion of these two structures can be found in \cite{Schiessel2} and here a short overview will be given.

For small $\beta \ll 1$ and $\alpha \approx \pi$ the chromatin fibers resemble solenoids where the linkers themselves follow closely a helical
path. This leads to the condition $\left ( \frac{d \cdot a}{R}\right ) \ll 1$. The lowest order approximation in $\left (\frac{d \cdot a}{R}
\right )$ indicates $b \approx d\sqrt{1 + a^2}$ (cf. Eq. \ref{eq:b^2}), $\alpha \approx \pi - \frac{a^2b}{R(1+a^2)}$ (cf. Eq. \ref{eq:alpha})
and $\beta \approx \frac{\pi-\alpha}{a}$ (cf. Eq. \ref{eq:beta}). These approximations lead to the geometrical properties of the solenoid-like
fibers: The radius of the fiber is $r$, $l_N$ is the length of a fiber of $N$ monomers, $ \lambda = N/l_N$ is the line density and $\varrho =
\lambda / \pi r^2$ is the density of the fiber:

\begin{align*}&r=\frac{b\;(\pi-\alpha)}{\beta^2+(\pi-\alpha)^2}\;\;\;\;\;\;\;\;
l_N=\frac{b\;N\;\beta}{\sqrt{\beta^2+(\pi-\alpha)^2}}\\
&\lambda =\frac{\sqrt{\beta^2+(\pi-\alpha)^2}}{b\;\beta }\;\;\;\; \varrho=\frac{(\beta^2+(\pi-\alpha)^2)^{5/2}}{\pi \; b^3 \;\beta
\;(\pi-\alpha)^2}\quad .\end{align*}

The vertical distance $\delta$ between two loops plays an important role in the following sections. It can be obtained by $\delta \approx
l_{\frac{2\pi}{\pi-\alpha}}$, which leads to
\begin{equation}\label{eq:delta_solenoid} \delta \approx \frac{ \left (\frac{2\pi}{\pi-\alpha} \right ) \; b \; \beta}
{\sqrt{\beta^2+(\pi-\alpha)^2}}\quad .
\end{equation}
One can calculate the exact value of $\delta$ by:
\begin{align*} \delta &= \overset{2 \pi r /a}{\underset{t=0}{\int}} m(\gamma(t))\cdot|(\gamma_x'(t),\gamma_y'(t))^t|  \; dt \\
&=\frac{4b\pi \sin(\frac{\alpha}{2})\sin(\frac{\beta}{2})}{\arccos\left(2\cos^2(\frac{\beta}{2})\sin^2(\frac{\alpha}{2})-1\right)
\sqrt{3+\cos(\alpha) + (\cos(\alpha)-1)\cos(\beta)} }
\end{align*}
However, the approximation (\ref{eq:delta_solenoid}) is much more useful.

For $\beta \ll (\pi-\alpha)$ one finds a very dense spiral ($ \delta \ll r$) and in the opposite limit $\beta \gg (\pi-\alpha)$ the solenoid has
a very open structure ($\delta \gg r$).

Next we turn to the crossed-linker fibers. Now consider the case $\beta \ll 1$ and $\alpha \ll \pi$: Above the regular polygons were discussed.
For a non-vanishing $\beta$ these regular polymers open up in an accordion-like manner. This leads to a three-dimensional fiber with crossed
linkers (cf. Fig. \ref{fig:phase_diagram}). Using $\beta \ll 1$ and Eq. \ref{eq:b^2}, Eq. \ref{eq:alpha} and Eq. \ref{eq:beta} one gets (cf.
\cite{Schiessel2}):
\begin{align*}
d^2=\frac{1}{4}\beta^2(4R^2-b^2)
\end{align*}
and thus
\begin{align*}
&r=\frac{b}{2\cos(\alpha/2)}\left (1-\frac{\beta^2}{4} \cot^2(\frac{\pi-\alpha}{2})\right ), \\
&l_N= \frac{N \;\beta \;b}{2}\cot(\frac{\pi-\alpha}{2}) \\
&\lambda = \frac{4}{\alpha \;\beta \; b},\quad  \varrho = \frac{16}{\pi \; \beta \; \alpha \; b^3} \quad .
\end{align*}
Now $\delta$ follows again from $\delta \approx l_{\frac{2\pi}{\pi-\alpha}}$:
\begin{equation}\label{eq:delta_crossed-linker}
\delta \approx \frac{1}{2}\left (\frac{2\pi}{\pi-\alpha} \right ) \;\beta \;b\tan\left (\frac{\alpha}{2} \right)\quad .
\end{equation}

Of course, not all points $\gamma_n \in \Gamma_n$ have the same probability to occur in real chromatin fibers. There are even some forbidden
areas in the phase space due to excluded volume restrictions. These forbidden areas are the subject of this section. In this numerical and
analytical study hard-core potentials were used to model the excluded volume interactions of the fibers. The border which separates the
non-excluded volume structures from those which fulfill the excluded volume condition is denoted by $\zeta$ (which is a function of $\alpha$ and
is plotted in Fig. \ref{fig:phase_diagram}, too).

\begin{figure}[ht]
\begin{center}
\includegraphics[width=\textwidth ,angle=0]{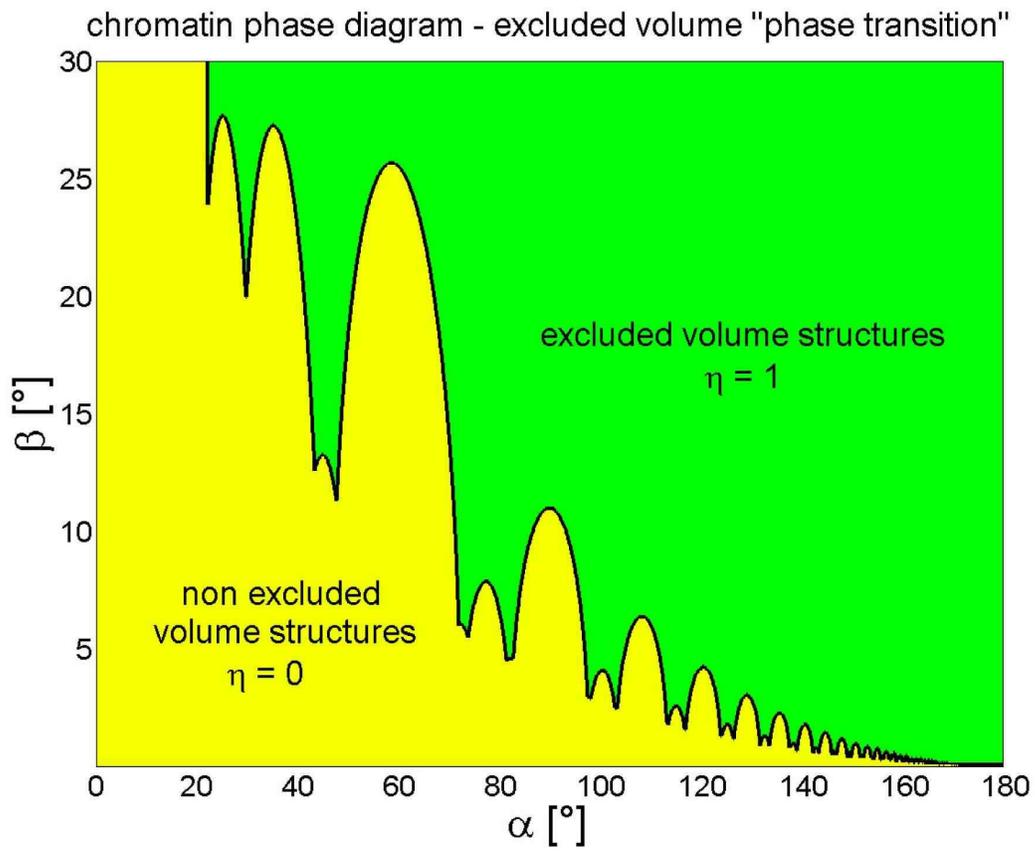}
\caption{\label{fig:excl_vol_main} Fine-structure of the excluded-volume "phase transition". The chromatin fibers in the green area fulfill the
excluded volume conditions, those in the yellow area do not. The borderline is the function $\zeta(\alpha)$.}
\end{center}
\end{figure}

$n_{min}=3 \left ( \frac{2\pi}{\pi-\alpha}-1\right)=3\left
(\frac{\pi+\alpha}{\pi-\alpha}\right )$ is given by the chromatin geometry (linker length, octamer diameter and height, see below). As shown in
Fig. \ref{fig:phase_diagram} the interesting part of the phase diagram for the excluded-volume phase-transition is the lower one with $\beta \in
[0,30°]$. This cut-out of the phase diagram is shown in Fig. \ref{fig:excl_vol_main}.

One can roughly explain Fig. \ref{fig:excl_vol_main} by dividing the excluded volume interactions between the nucleosomes in \emph{short-range}
and \emph{long-range} interactions: Between two consecutive nucleosomes there is never an excluded volume interaction, short-range interactions
occur between nucleosome $k$ and nucleosome $k+2$, and long-range interactions are those between nucleosomes with a larger distance than 2
($\Delta k>2$). These approximations can be found in \cite{Schiessel2}, too.

\begin{figure}[ht]
\begin{center}
\includegraphics[width=\textwidth ,angle=0]{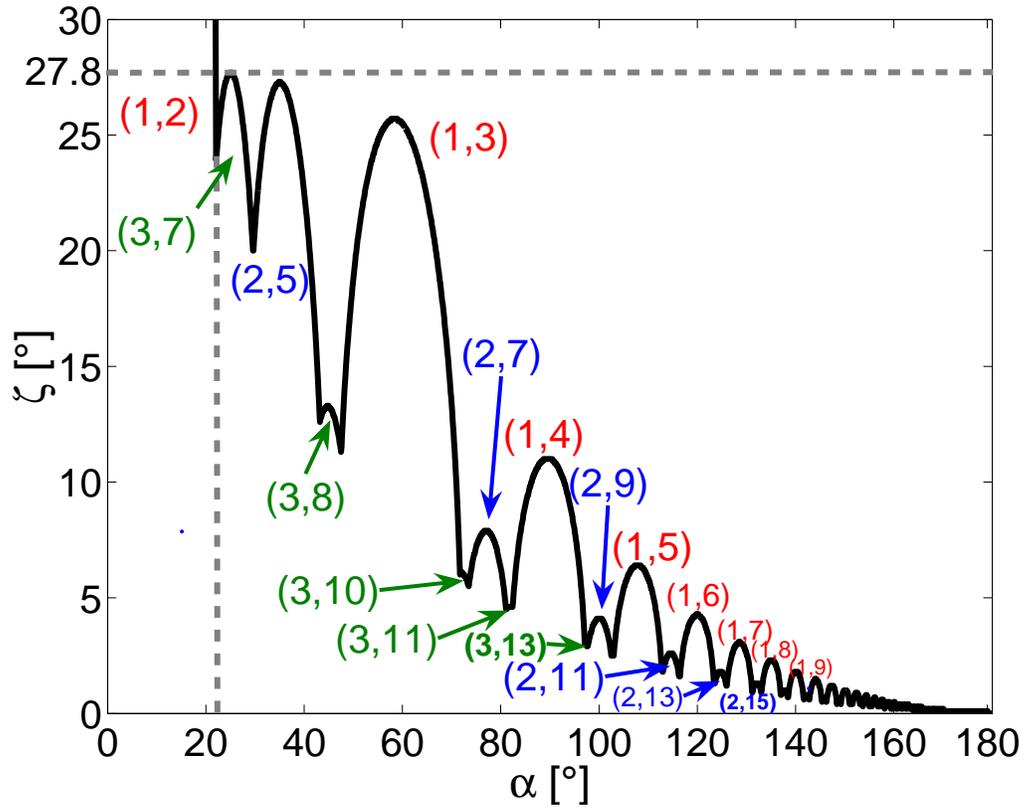}
\caption{\label{fig:fine_structure} Classification of the excluded volume regions.}
\end{center}
\end{figure}

Another explanation of the structure of the excluded-volume phase-transition, which even allows to derive the explicit behavior of $\zeta$ and
does not need to divide into short-range and long-range interactions, is presented below.$\\$
Every peak of the $\zeta$-function corresponds to a regular polygon: The large peaks
correspond to polygons of order 1 and the smaller ones to order 2 and 3 polygons. Between two order-1-peaks there is one order-2-peak and two
order-3-peaks. The classification of these $\zeta$-peaks is shown in Fig.~\ref{fig:fine_structure}: the $\zeta$-function has local maxima at every $\alpha_i^n$. The planar structures which belong to
$(\alpha_i^n,\beta=0)$ need a large rise of $\beta$ to arrive in the area of the excluded volume structures, because at $\beta=0$ nucleosome $k$
and $k+\frac{n \cdot 2\pi}{\pi-\alpha_i^n}$ are located exactly at the same position.$\\$ At first, consider only the special values
$\alpha_i^n$: At first order increasing $\beta$ from $\beta=0$ to $\beta=\tilde{\beta}$ leads to a vertical movement $\Delta_i^n(\tilde{\beta})$
of the relevant nucleosomes $k$ and $k+\frac{n \cdot 2\pi}{\pi-\alpha_i^n}$ along the chromatin axis . If $\Delta_i^n(\tilde{\beta}) = d \;
\Rightarrow \; \zeta(\alpha_i^n)=\tilde{\beta}$. For large $\alpha$, $\Delta_i^n(\tilde{\beta})$ can be calculated
 by Eq.\ \ref{eq:delta_solenoid} and one finds:
\begin{equation}\label{eq:Delta_i^n}\Delta_i^n(\tilde{\beta}) = \frac{ n \cdot\left (\frac{2\pi}{\pi-\alpha_i^n} \right ) \; b \; \tilde{\beta}}
{ \sqrt{\tilde{\beta}^2+(\pi-\alpha_i^n)^2}}\end{equation}with $\beta>0$ this leads to
\begin{equation}\label{eq:beta_von_delta}
\tilde{\beta}_i^n(\Delta_i^n)=\sqrt{\frac{(\Delta_i^n(\pi-\alpha_i^n))^2}{b^2 \; n^2\left ( \frac{2\pi}{\pi-\alpha_i^n}\right )^2-
(\Delta_i^n)^2}   }
\end{equation}
and furthermore $\tilde{\beta_i^n}(d)=\zeta(\alpha_i^n)$ implies
\begin{equation}\zeta(\alpha_i^n)=\sqrt{\frac{d^2 \cdot(\pi-\alpha_i^n)}{b^2 \; n^2\left ( \frac{2\pi}{\pi-\alpha_i^n}\right )^2-
d^2}}.\end{equation} Fig.\ \ref{fig:fine_structure} shows the numerically calculated zeta-function and the theoretical predictions for
the maxima. As Eq.\ \ref{eq:Delta_i^n} shows, $\Delta_i^n(\tilde{\beta}) \sim n$. This means that planar structures at an $\alpha_i^n$ of
\emph{higher} order need a \emph{smaller} rise of $\beta$ to fulfill the excluded volume conditions and therefore $\zeta$ is decreasing with
increasing $\alpha$. In fact this is the reason why $n$ is called the \emph{order} of the peaks of the $\zeta$-function. It can be understood
easily if one considers the fact that a higher order means more nucleosomes are located between two overlapping ones. And thus a rise of $\beta$
has a stronger effect than at lower orders.$\\$

As $\Delta_i^n(\tilde{\beta}) \rightarrow 0$ for $\alpha_i^n \rightarrow \pi$ the maxima of the $\zeta$-function converge towards 0 for
$\alpha_i^n \rightarrow \pi$. There are infinitely many $\alpha_i^n$ for $n \rightarrow \infty$ and the distance $\Delta_{i}^n$ converges to
zero, so the $\zeta$-function has infinitely many local maxima and minima for $\alpha_i^n \rightarrow \pi$. It is clear that $\zeta(\pi)=0$ as
the fiber forms a fully stretched fiber (a circle of radius $r=\infty$). This explains the forbidden strip at the left side of the figures as
the maximum of highest order $(1,2)$ with a corresponding one-dimensional structure.$\\$

So far the $\zeta$-function is only known at the positions of the maxima $\alpha_i^n$. Now consider values of $\alpha$, which are close to an
$\alpha_i^n$, say $\alpha' = \alpha_i^n \pm \Delta \alpha$ ("close" means such $\Delta\alpha$, which lead to a shift $\Delta a < 2r$). This
leads to a slight shift $\Delta a$ of those nucleosomes which are located at the same places (namely $k$ and $k+ \left ( \frac{n\cdot 2\pi}{\pi
- \alpha} \right )$). At $\beta=0$ this shift $\Delta a$ is orthogonal to the fiber's axis. This time $\Delta_i^n$ still denotes the distance
between the nucleosomes $k$ and $k+ \left ( \frac{n\cdot 2\pi}{\pi - \alpha} \right )$ along the fiber's axis, but now they are not located at
the same spots but slightly shifted. Therefore their distance $\Delta$ is no longer equal to their distance $\Delta_i^n$ along the axis, when
increasing $\beta$. $\\$ In this case increasing $\beta$ still leads to a movement along the vertical axis of the fiber, but now the distance
$\Delta$ of nucleosome $k$ and $k+ \left ( \frac{n\cdot 2\pi}{\pi - \alpha} \right )$ increases like $\Delta^2 \approx \Delta a^2 +
(\Delta_i^n)^2$. The fiber fulfills excluded volume, when $\Delta = 2r$, which means
\begin{equation*} \Delta_i^n = \sqrt{4r^2- \Delta a^2} < 2r. \end{equation*}
So the critical value of $\Delta_i^n$, which has to be achieved to fulfill excluded volume, decreases with increasing $\Delta a$. Therefore the
$\zeta$-function has a local maximum at $\alpha_i^n$: $\zeta(\alpha_i^n \pm \Delta \alpha) < \zeta(\alpha_i^n)$. $\\$

To calculate $\Delta a_{i,n}(\Delta \alpha)$ imagine a planar structure of $j$ nucleosomes with an entry-exit-angle $\alpha' = \alpha_i^n \pm
\Delta \alpha$ of two consecutive octamers. The locations of these nucleosomes are denoted by $p_0, \; p_1, \; p_2, \; ..., \; p_{j-1} \in
\mathbb{R}^3$. Without loss of generality one can assume
\begin{equation*} p_0 = \begin{pmatrix} 0 \\ 0 \\ 0 \end{pmatrix}, \; p_1= \begin{pmatrix} b&\cos(\frac{\alpha'}{2})\\ b&\sin(\frac{\alpha'}{2})\\
&\;\; 0 \end{pmatrix}\text{ and}\end{equation*}
\begin{equation*} \forall k \geq 2 \; : \; p_k = p_{k-1} + \mathcal{R}\cdot (p_{k-1}-p_{k-2})\end{equation*}
\begin{equation*}\text{with } \mathcal{R}:= \mathcal{R}_{\pi-\alpha'}^z=
\begin{pmatrix}
\cos(\pi-\alpha ')&  \sin(\pi-\alpha') \quad 0\\
-\sin(\pi-\alpha')& \cos(\pi-\alpha') \quad  0 \\
\quad 0 \quad &\quad \quad \quad \; \: 0  \quad \quad  1
\end{pmatrix}\end{equation*}
the rotational matrix along the $z$-axis. This leads to
\begin{equation*}p_{k}= \overset{k-1}{\underset{m=0}{\sum}} \mathcal{R}^m \cdot p_1 \quad \forall \; k >0.\end{equation*}
Now $\Delta a_{i,n}$ is given by $\Delta a_{i,n} = \|p_{k=i}\|$, which leads to
\begin{equation}\Delta a_{i,n}= \left \| \; \overset{i}{\underset{m=1}{\sum}} \mathcal{R}^{m-1} \cdot p_1 \; \right \|\end{equation}
(were $\mathcal{R} \text{ and } p_1 \text{ depend on }i, \; n \text{ and } \Delta \alpha$) Now one can use Eq.\ \ref{eq:beta_von_delta} to
calculate the $\zeta$-function around the maximal values $\alpha_i^n$:
\begin{equation*}\frac{ n \cdot\left (\frac{2\pi}{\pi-\alpha'} \right ) \; b \; \zeta(\alpha')}
{ \sqrt{\zeta(\alpha')^2+(\pi-\alpha')^2}} \overset{!}{=} \sqrt{4r^2- \Delta a^2} \end{equation*}
\begin{equation}\label{eq:zeta_for_large_alpha}
\overset{\zeta>0}{\Rightarrow} \; \zeta(\alpha'=\alpha_i^n \pm \Delta \alpha) =\sqrt{\frac{(4r^2-  \| \; \overset{i}{\underset{m=1}{\sum}}
\mathcal{R}^{m-1} \cdot p_1 \;  \|^2) (\pi-\alpha')^2}{b^2 \; n^2\left ( \frac{2\pi}{\pi-\alpha'}\right )^2-
4r^2+  \| \; \overset{i}{\underset{m=1}{\sum}} \mathcal{R}^{m-1} \cdot p_1 \;  \|^2}   }.
\end{equation}$\\$
Similar to the case of large $\alpha$ one can find equations for the $\zeta$-function of small $\alpha$: Eq.\ \ref{eq:delta_crossed-linker}
leads to
\begin{equation*}\Delta_i^n(\tilde{\beta})= \frac{1}{2}n \left (\frac{2\pi}{\pi-\alpha'} \right ) \; \tilde{\beta} \;  b \; \tan(\frac{\alpha'}{2})
\end{equation*}
with $\alpha' = \alpha_i^n \pm \Delta \alpha$. Now again $\zeta(\alpha')$ follows from $\tilde{\beta}(\Delta_i^n= \sqrt{4r^2-\Delta a_{i,n}
^2})=\zeta(\alpha')$:
\begin{equation}\label{eq:zeta_for_small_alpha}
\Rightarrow \; \zeta(\alpha'=\alpha_i^n \pm \Delta \alpha)=\frac{\cot(\frac{\alpha'}{2}) }{n \left (\frac{2\pi}{\pi-\alpha'} \right ) \cdot b }
\; \sqrt{4r^2 - \| \; \overset{i}{\underset{m=1}{\sum}}
\mathcal{R}^{m-1} \cdot p_1 \;  \|^2}.  
\end{equation}
$| \Delta\alpha | < c$ fulfilling $\Delta a (c) = 2r$ gives the interval of the allowed $\alpha$-values: $\alpha' \in [\alpha_i^n-c,\;
\alpha_i^n+c]$ for a certain peak ($n$,$i$). The predictions of Eqs.\ \ref{eq:zeta_for_large_alpha} and \ref{eq:zeta_for_small_alpha} are shown
in Fig.\ \ref{fig:fine_structure}, together with the simulation results.
\section{Excluded volume restrictions of the DNA linkers}\label{Sec:The Coulomb repulsion of the DNA linkers}
\begin{figure}[ht]
\begin{center}
\includegraphics[width=\textwidth ,angle=0]{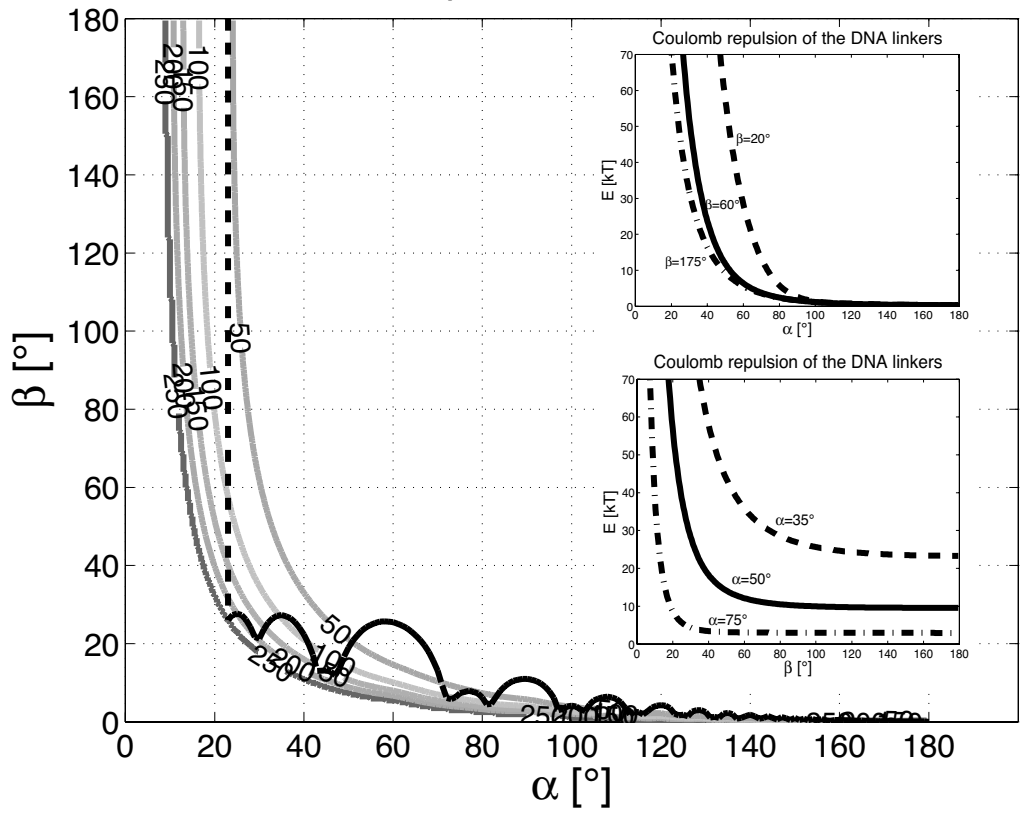}
\caption{\label{fig:equipotential} Coulomb repulsion of the DNA linkers. Shown are also two cuts in the inset.}
\end{center}
\end{figure}

As mentioned before the nucleosome-nucleosome excluded volume interactions are not the only ones within the chromatin fiber. The DNA linkers
have a diameter of about 2nm and therefore excluded volume restrictions, too. This is in particular very important for all crossed-linker
structures. But since the DNA linkers have a very strong (although screened) Coulomb repulsion their excluded volume interactions can be
revealed by looking at the Coulomb energy between the linkers. Other potentials like the nucleosome-nucleosome interaction or the interaction
between DNA linkers and nucleosomes will be neglected here. $\\$ One can use the Debye-H\"uckel theory to model the Coulomb repulsion of the DNA
linkers, but since the screening of this interaction starts at the radius of the DNA strand and due to the condensation of ions along the DNA
linkers, one has to calculate a correction of the screened potential by fitting the tail of the Debye-H\"uckel potential for an infinitely long
cylinder to the Gouy-Chapman potential in the far zone. This calculation can be found in \cite{SchellmanStigter,Stigter} and leads to a
corrected linear charge density $\nu_{\text{eff}}$ which is also given in the table below for different levels of monovalent salt concentration
and can be found in \cite{Schlick} for instance. During the last two to three decades DNA models based on such potentials have been developed
and applied widely, and their predictions are usually very good \cite{Allison,Klenin,Delrow,Klenin2,Merlitz,Hammermann,Fujimoto}.
\begin{center}\begin{tabular}{|c|c|c|c|c|c|}
\hline  $c_S$ [$10^{-2}$M] &1&2&3&4&5 \\ \hline $\kappa$ [nm$^{-1}$] &0.330&0.467 &0.572 & 0.660 &0.738 \\ \hline $\nu_{\text{eff}}$ [e/nm] &
-2.43&-2.96&-3.39&-3.91&-4.15 \\ \hline
\end{tabular}\end{center}
So one can calculate the Coulomb repulsion between the DNA segments $i$ and $j$ by evaulating
\begin{equation}
V_{i,j}=\frac{\nu_{\text{eff}}}{c}\int \int \frac{e^{-\kappa r_{i,j}}}{r_{i,j}} \text{d}x_i \text{d}x_j
\end{equation}
where $c$ is the total dielectric constant of water. These two integrals were numerically calculated and the results for the Coulomb energy of a
single chromatin linker within a fiber is shown in Fig.\ \ref{fig:equipotential}.

One can see that the Coulomb repulsion of the linkers is very high within the gaps of the excluded volume borderline. The repulsion also
diverges for the crossed linker fibers when $\beta$ becomes too small.

\section{Discussion and Conclusion}

The following effects have to be considered to explain the differences between the calculated theoretical
predictions and the simulational results.

The effective entry-exit angle $\alpha_{e}$ is the projection of $\alpha$ onto a plane which is orthogonal to the axis of the master solenoid.
It decreases with increasing $\beta$. This is a consequence of the fact that the length of the fiber increases with increasing $\beta$.
$\alpha_e$ is important for the calculation of the number of linkers which form a closed loop: $N_l= \frac{2\pi}{\pi-\alpha_e}$
($\frac{2\pi}{\pi-\alpha_e}$ gives the number of breaks of $\alpha_e$ which has to be done to get a full loop. If $\frac{2\pi}{\pi-\alpha_e}$ is
an integer, this is equal to the number of linkers, which corresponds to a full loop. If it is not an integer number, then the fractional part
gives the fraction of $\pi-\alpha_e$ which is missing for a full loop, but the entry-exit angle is fixed here and it is assumed that
$N_l=\frac{2\pi}{\pi-\alpha_e}$). In the calculations above $\alpha_e=\alpha=$const. was assumed, because only small $\beta$ were considered.
$\alpha_e$ will be calculated as a function of $\alpha$ and $\beta$ below. It converges towards $0$ for $\beta \rightarrow \pi$ for different
values of $\alpha$ and $\beta$.$\\$ To calculate $\alpha_e$ consider three consecutive nucleosome locations within a given fiber: $n_0$, $n_1$
and $n_2$ $\in \mathbb{R}^3$. Without loss of generality one can assume: $n_0 = p_0$, $n_1=p_1+p_0$ and $n_2=p_2+p_1+p_0$ were the $p_i$ are the
following linker vectors:
\begin{equation*}p_0 = \begin{pmatrix} 0 \\ 0 \\ 0\end{pmatrix} \text{, } p_1 = \begin{pmatrix} \sqrt{b^2-d^2} \\ 0 \\ d \end{pmatrix}
\text{ and }p_2 = \begin{pmatrix} x_2 \\ y_2 \\ d \end{pmatrix}\quad .\end{equation*} This means that the $z$-axis of the coordinate
system is the axis of the fiber ($d$ is given by Eq. \ref{eq:d of the spiral} and $b$ is the linker length). Now $\| p_2 \| = b$ leads to
\begin{equation*}y_2=\sqrt{b^2-d^2-x_2^2}\end{equation*}

and $x_2$ can be calculated by using
\begin{equation*}\cos(\alpha)=\frac{\langle-p_1 \mid p_2 \rangle}{b^2}\end{equation*}
\begin{equation*}\Rightarrow x_2 = -\frac{b^2\cos(\alpha)+d^2}{\sqrt{b^2-d^2}} \end{equation*}
$\text{ with }d=\frac{b \sin \left ( \frac{\beta}{2} \right )}{\sqrt{\csc^2 \left (\frac{\alpha}{2} \right ) - \cos^2 \left ( \frac{\beta}{2}
\right ) } }\text{ this leads to }$
\begin{equation*}y_2=\sqrt{b^2-d^2-\frac{(b^2\cos(\alpha)+d^2)^2}{b^2-d^2}}\overset{d}{=}
\sqrt{b^2\cos^2\left(\frac{\beta}{2}\right)\sin^2(\alpha)}\end{equation*}

and now $\alpha_e$ follows from the projection onto the $x$-$y$-plane:
\begin{equation*}\frac{\sin(\alpha_e)}{2}=\frac{\sqrt{(x_1+x_2)^2+y_2^2}}{2\sqrt{x_2^2+y_2^2}}\end{equation*}
\begin{equation} \label{eq:alpha_eff} \Rightarrow \;\sin \left ( \frac{\alpha_e}{2} \right ) = \sqrt{\frac{\cos\left(\frac{\beta}{2}\right)
\sin^2\left(\frac{\alpha}{2}\right)}{2(1+\cos(\alpha))}} \quad .\end{equation} As Eq. \ref{eq:alpha_eff} shows, $\alpha_e$ decreases with
increasing $\beta$. As a consequence of this $N_l$ decreases, too: If $\alpha_e \rightarrow 0$, then
$N_l(\alpha,\beta)=\frac{2\pi}{\pi-\alpha_e(\alpha,\beta)} \rightarrow 2$. So all fibers with high values of beta need only approx. 2
nucleosomes for a complete loop. This is confirmed by Fig. \ref{fig:phase_diagram}. As $\alpha_e$ was assumed to be constant, $N_l$ was also
constant in the calculations above. This is a suitable approximation for small $\beta$. For larger $\beta$ the assumed values of $N_l$ were a
bit too large and therefore the calculated values of $\zeta$ were a bit too small - but this effect is small compared to the other estimations
which were made above. The error of $\zeta$, due to the assumption that $N_l$ is constant, increases with increasing $\beta$. $\\$ So to
increase accuracy in the calculations above one should replace the term $\frac{2\pi}{\pi-\alpha'}$ by
$\frac{2\pi}{\pi-\alpha_{e}(\alpha',\beta)}$ and furthermore one should use Eq. \ref{eq:d of the spiral} to derive exact equations in order to
replace Eqs. \ref{eq:delta_solenoid} and \ref{eq:delta_crossed-linker}. But this can only be done by using numerical methods and not
analytical ones. Moreover, Fig. \ref{fig:fine_structure} shows that the assumed approximations lead to useful equations, which describe
$\zeta$ in a suitable way.

\section*{Acknowledgments}

We would like to thank J\"org Langowski, Jens Odenheimer and Frank Aumann for the fruitful discussions.

\vfill\eject

\vfill\eject

\vfill\eject

\vfill\eject
\begin{figure}[ht]
\begin{center}
\includegraphics[width=\textwidth ,angle=0]{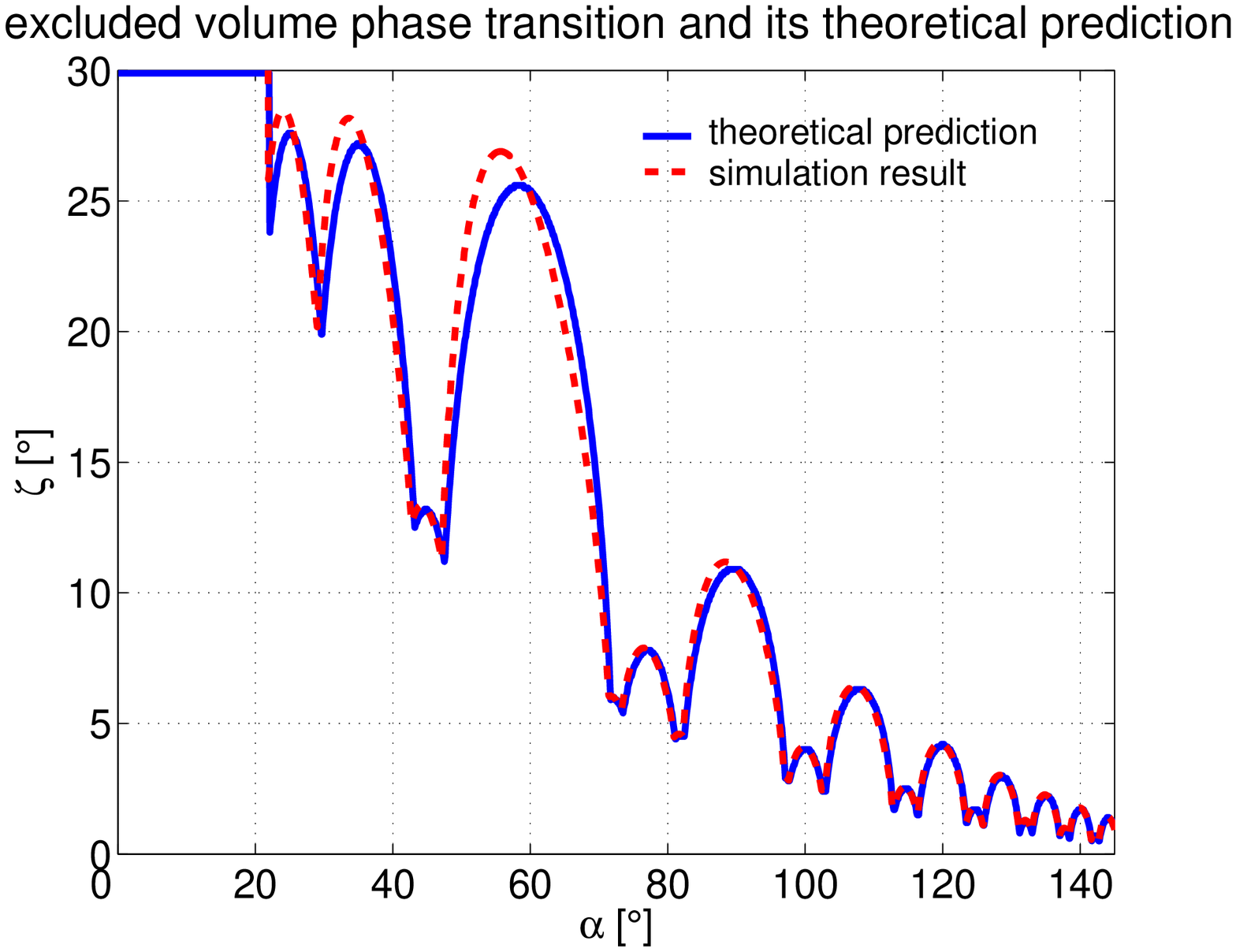}
\caption{\label{fig:result1} Calculated theoretical prediction (red) and the simulation result (blue) for $\zeta$ and small $\alpha$.}
\end{center}
\end{figure}

\vfill\eject
\begin{figure}[ht]
\begin{center}
\includegraphics[width=\textwidth ,angle=0]{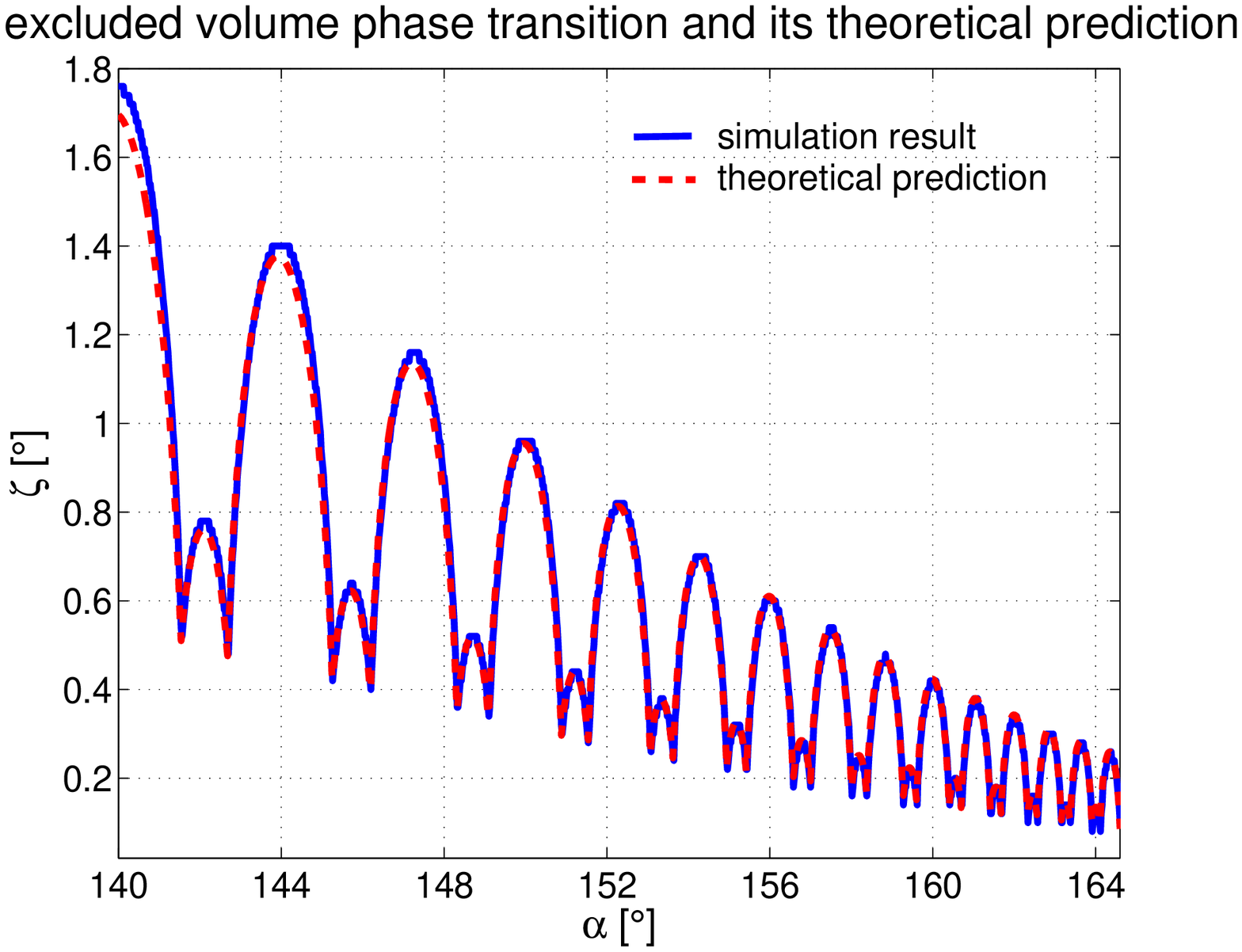}
\caption{\label{fig:result2} Calculated theoretical prediction (red) and the simulation result (blue) for $\zeta$ and larger $\alpha$.}
\end{center}
\end{figure}

\vfill\eject

\vfill\eject

\end{document}